\documentclass[a4paper,11pt]{article}
\pdfoutput=1 
\usepackage{siunitx}
\usepackage{jinstpub} 

\title{\boldmath Evaluation of Radiation Hardness of High-Voltage Silicon  Vertical JFETs}

\author[a,1]{Gabriele Giacomini,\note{Corresponding author.}}
\author[b]{Marco Bomben}
\author[a]{Wei Chen,}
\author[a]{David Lynn,}

\affiliation[a]{Brookhaven National Laboratory, Upton, 11973, NY, USA}
\affiliation[b]{LPNHE, 4 Place Jussieu,  75005 Paris, France}

\emailAdd{giacomini@bnl.gov}

\abstract{In the future ATLAS Inner Tracker, each silicon strip module will be equipped with a switch able to separate the high voltage supply from the sensor in case the latter becomes faulty. The switch, placed in between the HV supply and the sensor, needs to sustain a high voltage in its OFF state, to offer a low resistance path for the sensor leakage current in the ON state, and be radiation hard up to $1.2\cdot 10^{15}~ {\rm n_{eq}/cm^2}$ along with other requirements. While GaN JFETs have been selected as suitable rad-hard switch, a silicon vertical HV-JFET was developed by Brookhaven National Laboratory as an alternative option. Pre-irradiation results showed the functionality of the device and proved that the silicon HV-JFET satisfied the pre-irradiation requirements for the switch. To assess its suitability after irradiation, a few $p$-type HV-JFETs have been neutron irradiated at Jozef Stefan Institute (JSI, Ljubljana, Slovenia). This paper reports the static characterization of these irradiated devices  and the TCAD numerical simulations  used to get an insight of the physics governing the post-irradiation behaviour. }


\begin{document}
\maketitle
\flushbottom
\section{Introduction}
\label{sec:intro}

At the High-Luminosity Large Hadron Collider (HL-LHC) at CERN (Geneve, Switzerland), the ATLAS experiment will run with a completely renovated Inner Tracker (ITk) with respect to its LHC phase~\cite{a}. One of the differences in this upgrade is that currently either three or four silicon microstrip sensors will be biased by a single common High-Voltage (HV) line. If one of these three or four sensors  fails and starts drawing a high current, all the other good detectors connected to the same HV line must be switched off, thus losing a large detector area. To avoid this situation the faulty detector must be disconnected from the common HV line. The solution is to place a switch between each strip sensor and the common HV line. Normally the switch is closed (ON State) which means that a little voltage falls on it and almost the full voltage is applied to the sensor while the sensor leakage current flows through the switch. If the sensor gets damaged then the switch is opened (OFF state) and the full voltage falls on the switch. There is little or no current from the faulty sensor which is kept at about zero voltage but the remaining good  devices connected to the same HV line continue working normally. The requirements for such a switch so that it can be usable in the ITk are:
\begin{itemize} 
\item in the OFF state, capable of operating above 600~V (with an equivalent source-drain resistance ${\rm R_{ds, OFF}}>100~ {\rm M\si{\ohm}}$);
\item in the ON state,  having a ${\rm R_{ds, ON}}  < 10-100~ {\rm \si{\ohm}}$;
\item capable of operating in a 2T magnetic field;
\item survive radiation doses of 50 Mrad and fluences of $1.2\cdot10^{15} ~{\rm n_{eq}/cm^2}$.
\end{itemize} 

These requirements exclude, among others, electro-mechanical switches (which suffer in high magnetic fields), power MOSFETs (since their threshold voltage ${\rm V_{th}}$ changes with irradiation) and power transistors (their gain changes with irradiation).
A program, called HV-Mux was initiated to seek after such a switch ~\cite{b}. Silicon carbide and gallium nitride JFETs have been extensively investigated by means of irradiation campaigns. The latter satisfies all the specifications listed above and has been selected  be used as the switch for the ITk.
However, prior to establishing its suitability, we at Brookhaven National Laboratory (BNL) we developed a new kind of HV silicon JFET which satisfies most of these requirements before being irradiated. To assess its suitability as a rad-hard switch at the fluence expected at the ITk, some $p$-type samples were neutron irradiated at the TRIGA nuclear reactor run by the Jozef Stefan Institute (JSI, Ljubljana, Slovenia). The fluences at which the JFETs were exposed were: $4 \cdot 10^{14}$, $8 \cdot 10^{14}$ and $1.5 \cdot 10^{15}~ {\rm 1~ MeV~ n_{eq}/cm^2}$.

This paper is organized as follows: a brief description of the device and its functioning is presented in Section~2. The results, consisting in a static characterization at the probe station, are shown  in Section~3.  TCAD numerical simulations that have been performed to get an insight of the physics governing the irradiated devices are  presented in Section~4.

\begin{figure}
\centering 
\includegraphics[width=.5\textwidth]{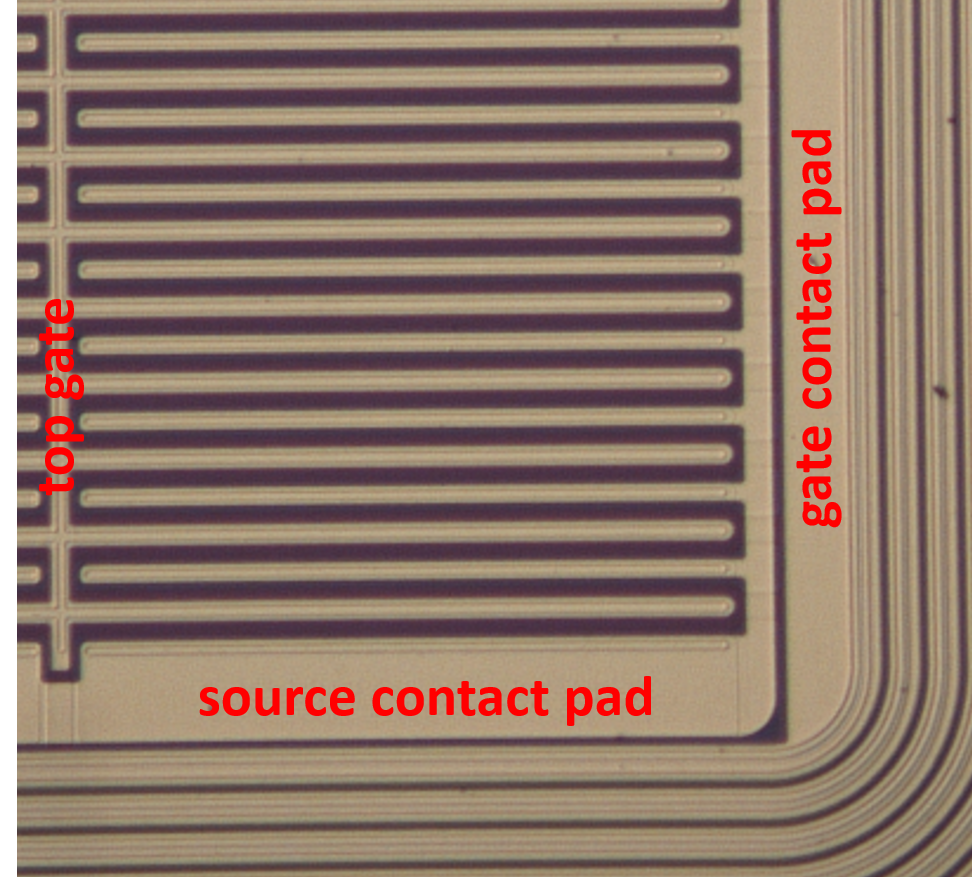}
\caption{\label{fig:photo} Microscope picture of a quarter of a HV-JFET, showing the positions of the gate and source in the device, while the drain is the back contact itself. The picture shows an area approximately 0.5~mm~x~0.5~mm large.}
\end{figure}

\begin{figure}
\centering 
\includegraphics[width=.9\textwidth]{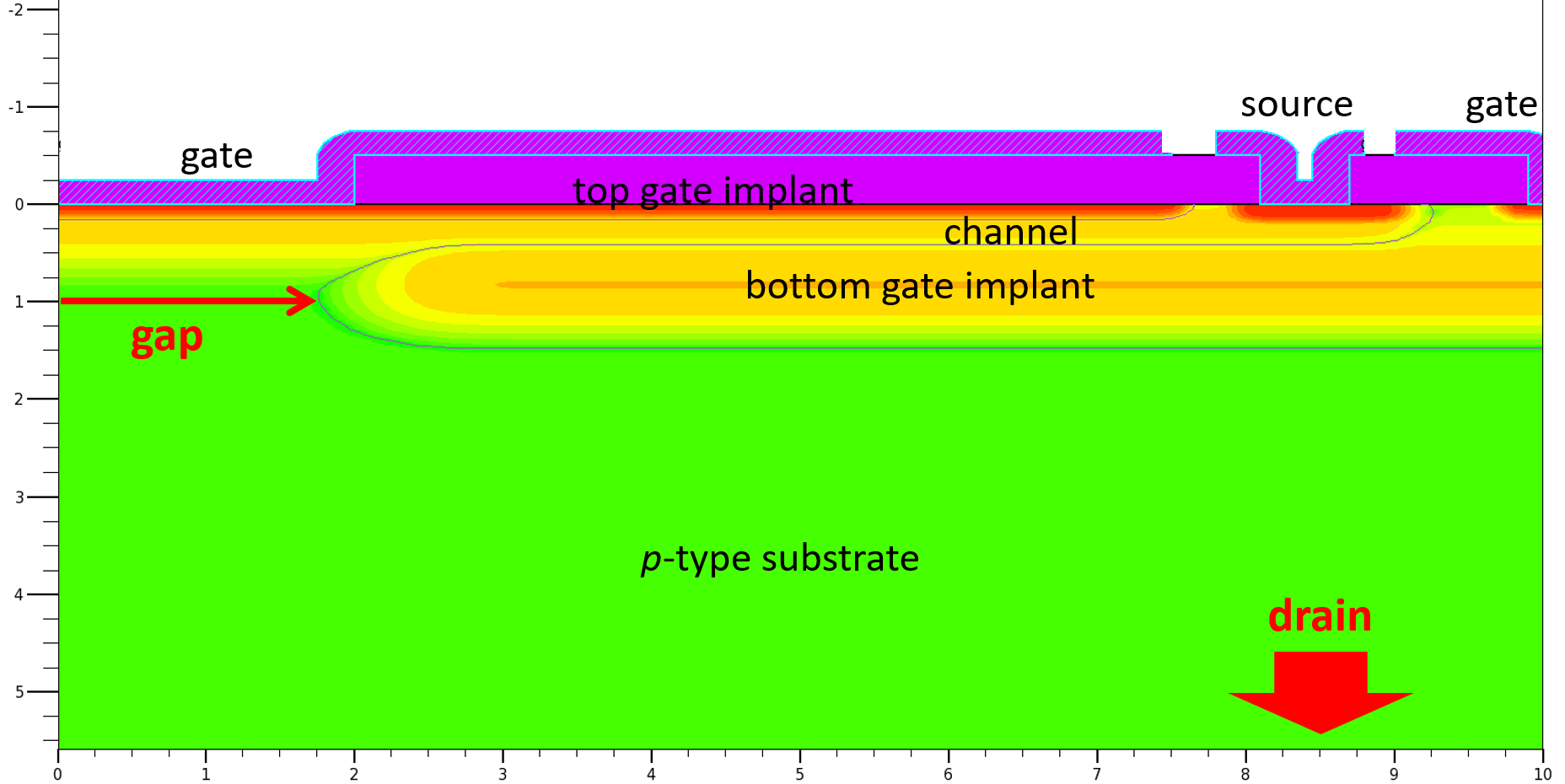}
\caption{\label{fig:geometry} Cross sectional view of a typical HV-JFET. This geometry has been used in all the TCAD simulations of Section 4. The structure is symmetric with respect to the vertical axis X=0.}
\end{figure}

\section{The HV silicon JFET}
\label{sec2}

Figure~\ref{fig:photo} shows a microscope picture of a typical HV-JFET used in this study. It has been fabricated on 4", $p$-type epitaxial wafers. The thickness of the epitaxial layer is 50~${\rm \mu m}$, which has been grown over a  thick handling wafer. The doping concentration of the epitaxial layer is $7\cdot 10^{13}~ {\rm cm^{-3}}$ resulting in a full depletion voltage of 130~V as measured on large-area diodes present as test structures on the wafer. We  expect a decrease of the substrate full depletion voltage after irradiation, even after the highest fluence used in this study ($1.5 \cdot 10^{15}~ {\rm 1~ MeV~ n_{eq}/cm^2}$): as reported in ~\cite{l}, in fact,  we expect a decrease of the effective doping concentration, due to acceptor removal (boron) for all the fluences used.  Figure~\ref{fig:geometry} shows the cross section of a simplified device, as used for all the TCAD numerical simulations of Section 4.  On the top side of the device, the source surrounds the top gate.  In the actual devices fabricated in our clean room, to increase the width of the device, and proportionally its current in the ON state, the source and the top gate are designed in an interdigitated configuration: the source  fingers  are shorted together by the metal layer and so are the top gate fingers. The total width of the device is about 20~cm.  The bottom gate runs all over the active area (about 1~mm~x~1~mm in the production) but features a gap of a few microns just below the middle of the top gate. The top gate is shorted through a deep  implant with the bottom gate, so that the JFETs are in a triode configuration, but this results in it not being possible to independently bias the two gates in this first prototyping fabrication. As in a  usual JFET, the channel runs all over the surface as well, connecting all the source implants together. Its length is defined by the overlap between the gates. The drain is the silicon substrate, contacted at the back of the wafer, making the HV-JFET a vertical structure, as is typical for silicon power devices~\cite{f}.   

The source-to-drain current flows between these two terminals passing through the gap in the bottom gate. As in a standard JFET, the gate voltage modulates the channel resistance. The device enters into saturation mode when the channel end is pinched off by the reverse bias applied between the channel itself, the gates and the drain. After that point, the current drifts to the drain. In the HV-JFET, the drain is far away from the channel and the channel is shielded from the drain voltage by the bottom gate. This means the drain voltage has little influence in closing the channel or, stated otherwise, very high drain voltages must be applied to significantly affect  the electrostatics at the channel end. This has been clearly shown by TCAD numerical simulations~\cite{d} and is of paramount importance when dealing with irradiated substrates.

We fabricated the $p$-type JFETs in a  batch of 4 wafers, each wafer differing in the implantation dose of the $p$-channel only. JFETs belonging to the two wafers with the lowest boron doses have the channel pinched-off already at $V_{gate}=0 ~V$. JFETs from the other two wafers are working and show different pinch-off voltage. Ultimately we irradiated the JFETs with the higher pinch-off voltage as to have a larger source current in the ON state.

The HV-JFETs have been entirely fabricated using  a standard planar process in the silicon processing facility of the Instrumentation Division at BNL~\cite{e}; another effort made by CNM (Barcelona, Spain) has fabricated  vertical silicon HV-JFETs using a 3D technology~\cite{c}.
The planar technology used at BNL is sub-optimal: as pointed out in~\cite{e}, a process involving a thin  epitaxial layer deposition after the bottom gate implantation avoids some undesired effects ("horn" effect), where the tails of the deep bottom gate compensate for the channel implant at the gap edges.  However, this fabrication must be seen as a proof of concept of the device and  improved performance  can be expected in an optimized fabrication.

\begin{figure}
\centering 
\includegraphics[width=.45\textwidth]{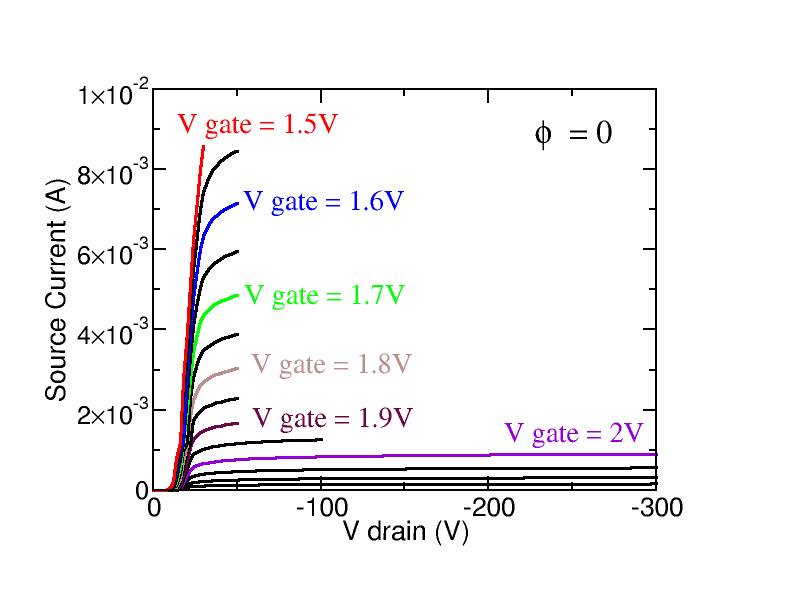}
\includegraphics[width=.45\textwidth]{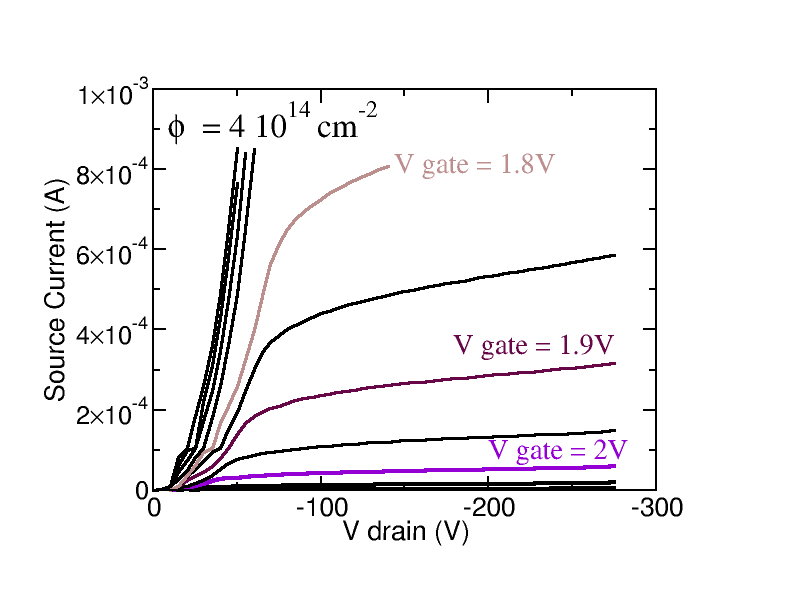}
\includegraphics[width=.45\textwidth]{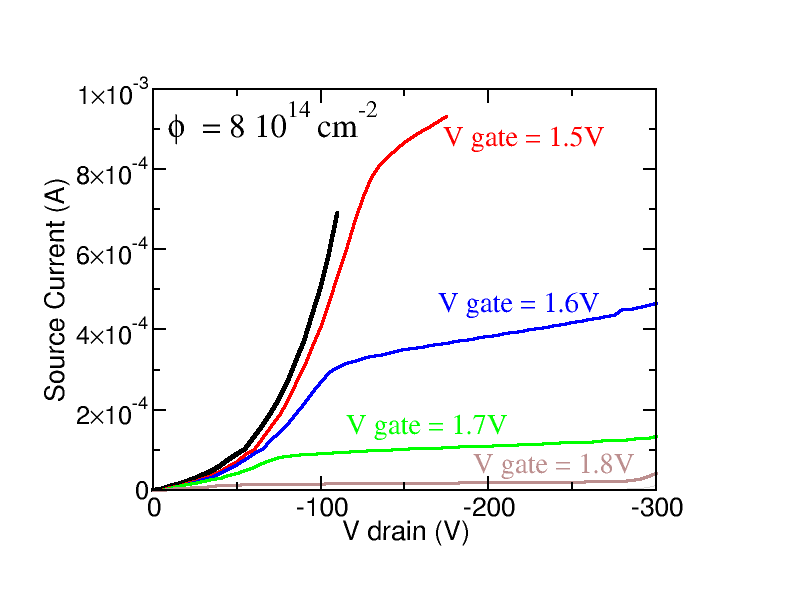}
\includegraphics[width=.45\textwidth]{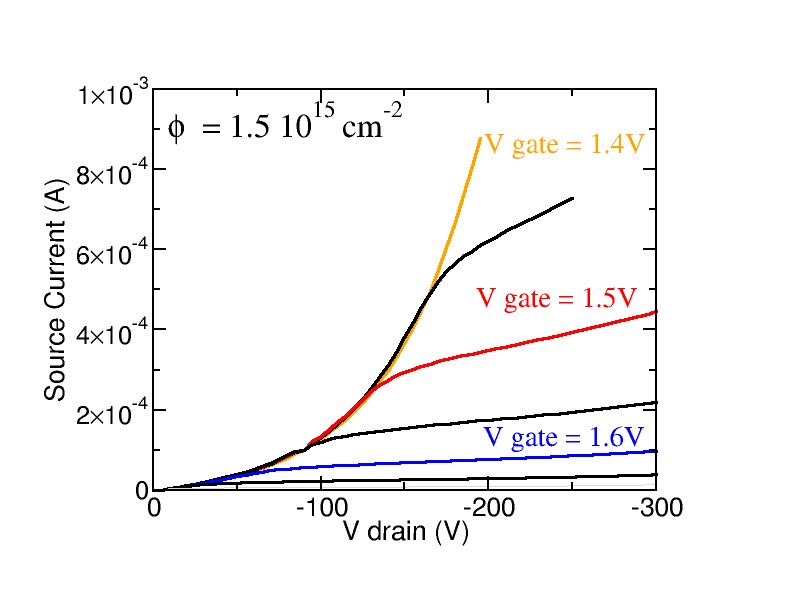}
\qquad
\caption{\label{fig:measure} Measurements of the output characteristics of HV-JFETs for the different fluences, from 0 to $ 1.5 \cdot 10^{15}~ {\rm 1~MeV~ n_{eq}/cm^2}$. Curves acquired with the same gate voltage have the same colour.}
\end{figure}

\begin{figure}
\centering 
\includegraphics[width=.6\textwidth]{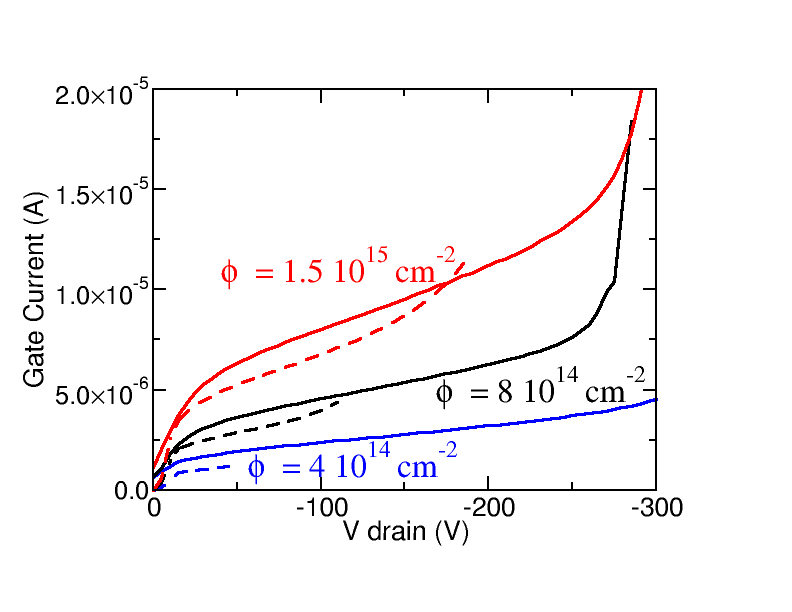}
\caption{\label{fig:gatecurrent} Measurements of the gate current for typical HV-JFETs irradiated at different fluences. Solid lines are for ${\rm V_{gate}=2~V}$, dash lines are for ${\rm V_{gate}=0~V}$.}
\end{figure}

\section{Electrical Characterization}
\label{sec3}
Several $p$-type HV-JFETs were irradiated at the TRIGA nuclear reactor of JSI, Ljubljana, Slovenia. The chosen irradiation fluences were $4 \cdot 10^{14}$, $8 \cdot 10^{14}$ and $1.5 \cdot 10^{15}~ {\rm 1 MeV ~n_{eq}/cm^2}$. 
Typical curves for the three irradiation levels as compared to the pre-irradiation case are shown in Figure~\ref{fig:measure}. To measure the output characteristics, the drain voltage is swept from 0~V to the breakdown voltage (which happens between gate and drain) and the currents at the source and the gate are measured. The drain voltage is applied through the chuck of a probe station and supplied by a Keithley 2410, while the gate and the source are contacted  by needles mounted on the probe station micro-manipulators; the current is then read-out by an HP4145B. However, to limit the power dissipation within the JFET, we had to limit the drain voltage for the highest currents (i.e. at the lowest gate voltages). In particular, due to the unfortunate absence of smaller devices, current-voltage characteristics for zero gate voltage are not measurable at high voltage where the device is in saturation.

The breakdown voltage is independent from the irradiation level: it happens for ${\rm V_{drain} =300~V}$, before and after irradiation. TCAD studies~\cite{d} show that the region with the highest electric field, where   the breakdown likely originates,  is at mid gap at the channel-top gate junction. The termination region, made by 16 floating guard rings surrounding the bottom gate has been measured to be able to sustain voltages in excess of 600~V, pre-irradiation. Irradiation is expected to increase the voltage handling capability of such termination, however, a direct measurement is not possible since the active region of the JFET breaks down much earlier.

For the same gate voltage, the source-to-drain current in saturation decreases with irradiation. Stated differently, the irradiation shifts the pinch-off voltage at lower gate voltages. This is due to the radiation-induced acceptor removal~\cite{l} in the channel, that increases the resistance of the same and therefore decreases the current.

Of particular interest is the shift of the drain saturation voltage by increasing the irradiation fluence. As in a standard JFET, as well as in the linear region of a  HV-JFET  (where the channel end is not yet pinched-off by the combined action of the gates and the drain),  the I-V curves lay together on the same envelope, for any gate voltage. This is evident for the highest irradiation fluence, while it holds with good approximation for the lowest.  The shape of this envelope is a strong function of the irradiation fluence, featuring a lower concavity at the higher fluences. As in the standard JFET, the drain supplies at the  end of the channel the voltage  that,  summed up with the gate voltage, closes the channel. At the fluences used in this work, the channel, which is relatively highly doped, experiences an acceptor removal~\cite{l} and one expects  that lower drain voltages are needed to close the channel. This clearly does not happen: higher drain voltages need  to be applied as the irradiation level increases. This points to the fact that other phenomena are at play: the physical interpretation that we propose to explain this behaviour is the subject of section 4.

Figure~\ref{fig:gatecurrent} reports the gate leakage currents for three HV-JFETs irradiated at different fluences. For clarity sake, only gate currents for ${\rm V_{gate} = 2~V}$ and ${\rm V_{gate} = 0~V}$ are reported, the other curves at intermediate voltages lying in between. We see a slight  dependence from the gate voltage, possibly due to interface effects at the external border of the device. At full depletion voltage, we expect a current equal  to: ${\rm I_{leakage}= V \cdot \alpha \cdot \phi}$, where $V$ is the volume (1~mm~ x~ 1~mm~ x~ 50~$\mu$m), $\alpha$  is the damage constant (having a value of $\alpha = 4 \cdot 10^{-17}~ {\rm  A/cm}$) and $\phi$ is the fluence (${\rm n_{eq}/cm^2}$). As can be seen in Figure~\ref{fig:gatecurrent},  gate currents hardly  saturate at the depletion voltage, making the extraction of the damage constant not realistic. However, taking for ${\rm I_{leakage}}$ the gate current at full depletion, a maximum value for the damage constant slightly larger than a factor of two than the one reported in the literature can be extracted.

If  such a  JFET  had to be used as a  switch  after a fluence of $1.5 \cdot 10^{15}$~ ${\rm 1~MeV~ n_{eq}/cm^2}$,  a  leakage current of a few mA (which is the room temperature leakage current of one $cm^2$ large silicon sensor 200-300 $\mu$m thick)  would flow from source to drain only for ${\rm V_{drain} > 200~V}$  in the ON state, resulting in a very high power dissipation of about 1~W within the device. Higher voltages must be applied by the voltage supply too, making all the system too impractical to be used. To enter the OFF state, the JFET can be closed by a low gate voltage (2V) and the current it draws is acceptable (consisting of the  leakage current generated within the depleted - but irradiated - substrate, order of 10 $\mu$A). However, for this particular JFET, the breakdown voltage does not satisfy the HV-Mux requirements.

\begin{table}[h]
\centering
\caption{\label{table} Trap parameters.}
\smallskip
\begin{tabular}{|c|c|c|c|c|}
\hline
trap type & energy level (eV) & density (${\rm cm^{-3}}$)& $\sigma_n$ (${\rm cm^{-2}}$)& $\sigma_p$ (${\rm cm^{-2}}$)\\
\hline
acceptor &      ${\rm E_c - 0.42}$   & $1.6 \cdot \phi$  &  $2 \cdot 10^{-15}$ &   $2 \cdot 10^{-14}$\\
acceptor &      ${\rm E_c - 0.46}$   & $0.9 \cdot \phi$   &  $5 \cdot 10^{-15}$ &   $5 \cdot 10^{-14}$ \\
donor    &      ${\rm E_v + 0.36}$   & $0.9 \cdot \phi$   &  $2.5 \cdot 10^{-14}$ & $2.5 \cdot 10^{-15}$ \\
\hline
\end{tabular}
\end{table}

\section{TCAD Numerical Simulations}
\label{sec4}
\begin{figure}[h]
\centering 
\includegraphics[width=.45\textwidth]{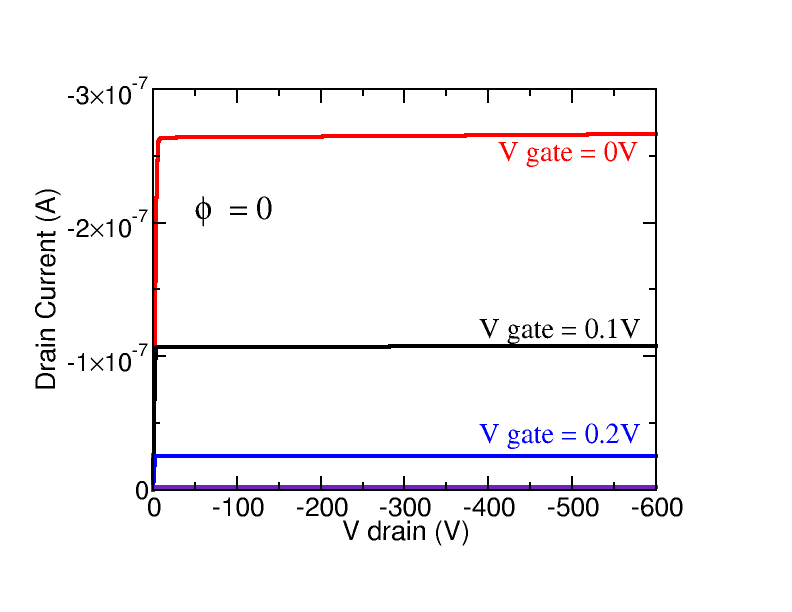}
\includegraphics[width=.45\textwidth]{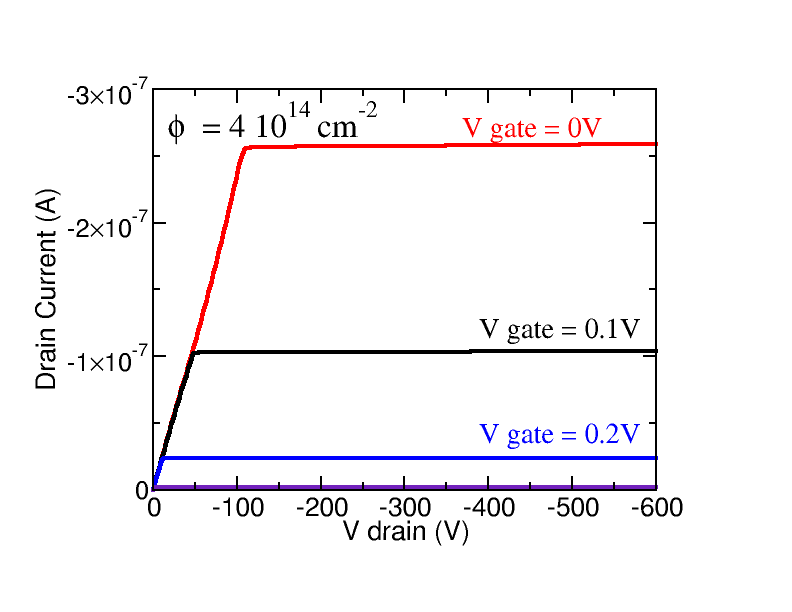}
\includegraphics[width=.45\textwidth]{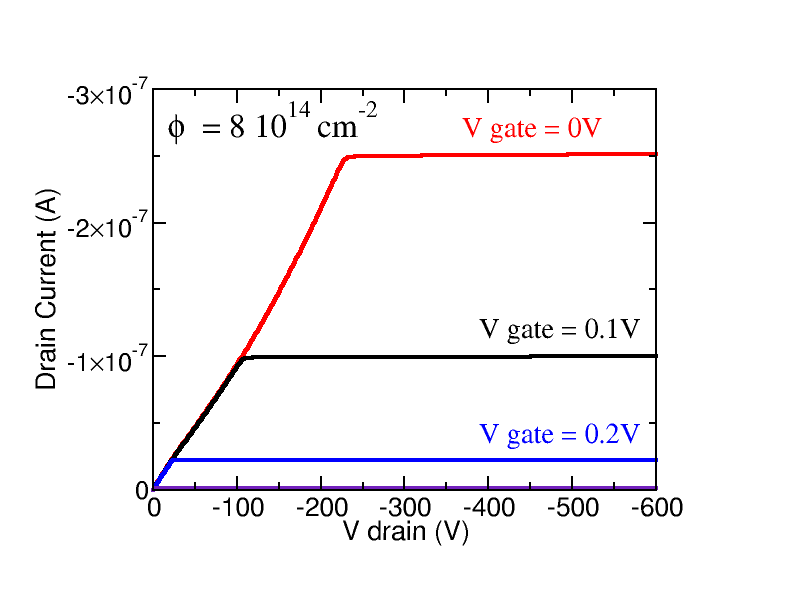}
\includegraphics[width=.45\textwidth]{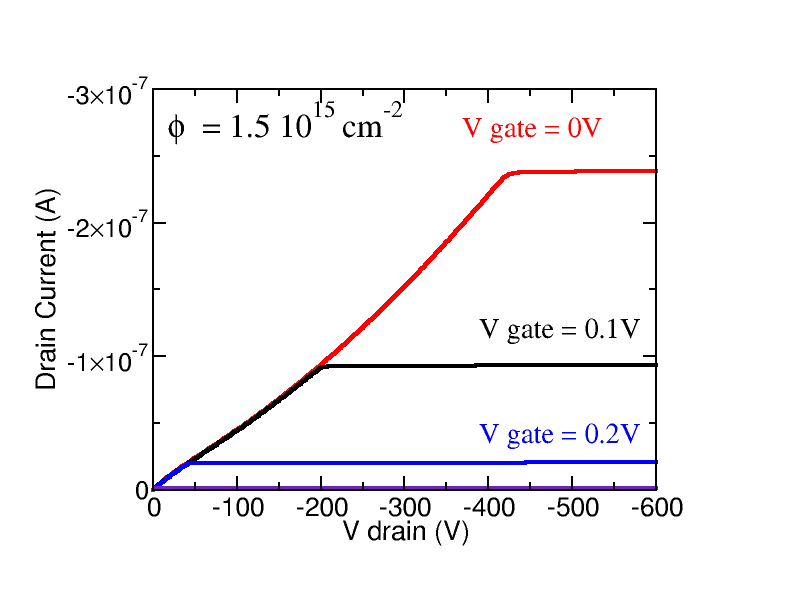}
\qquad
\caption{\label{fig:simulations} Simulated output characteristics for different fluences using the geometry of Figure~\ref{fig:geometry}.}
\end{figure}

Figure ~\ref{fig:simulations} shows TCAD simulations of a HV-JFET having the geometry of Figure~ \ref{fig:geometry}. The SILVACO simulator has been used~\cite{silvaco}. Standard models (without any impact ionization model activated) have been used. Simulations of the radiation damage are obtained  using three traps  as described by the "Perugia" model~\cite{g},   as shown in Table~\ref{table}.
In the table, the "energy level" is the energy gap in eV from the conduction or the valence band for an acceptor or donor trap, respectively;  "density" is the density of a trap in ${\rm cm^{-3}}$, $\phi$ is the fluence,  ${\rm \sigma_n (p)}$ is the capture cross section for electrons (holes). The Perugia model does not simulate the acceptor removal.

By changing the density of the traps according to the fluence, the I-Vs of Figure~\ref{fig:simulations} are obtained. As experienced in the measurements of real devices,  TCAD simulations also predict increasing drain voltages at the onset of saturation by increasing the fluence. This is due to the build-up of positive charge along the path of the hole current which is created by the ionized donor traps. A donor  trap is neutral when filled with an electron, while positively charged when empty (thus the name "donor"  traps): there are no electrons to fill up such traps and the holes of the source-to-drain current rapidly recombine with the electrons of the filled traps. The net-positive charge creates potential barriers in the gap region that inhibit the injection of holes and block the drain voltage from closing the channel end. This effect is more severe at higher irradiation levels, since more traps are generated. Acceptor traps, on the other hand, are neutral because they are not filled by electrons, and being neutral do not alter the electrostatics of the device. Therefore, larger drain voltages must be applied to lower the potential barrier and provide the voltage sufficient for the pinch-off of the channel end. 

Substrate effects of increased substrate resistance which quenches the current by providing a resistive drop are excluded, since the resistivity saturates at these irradiation fluences~\cite{h}~\cite{i}.

\begin{figure}
\centering 
\includegraphics[width=.9\textwidth]{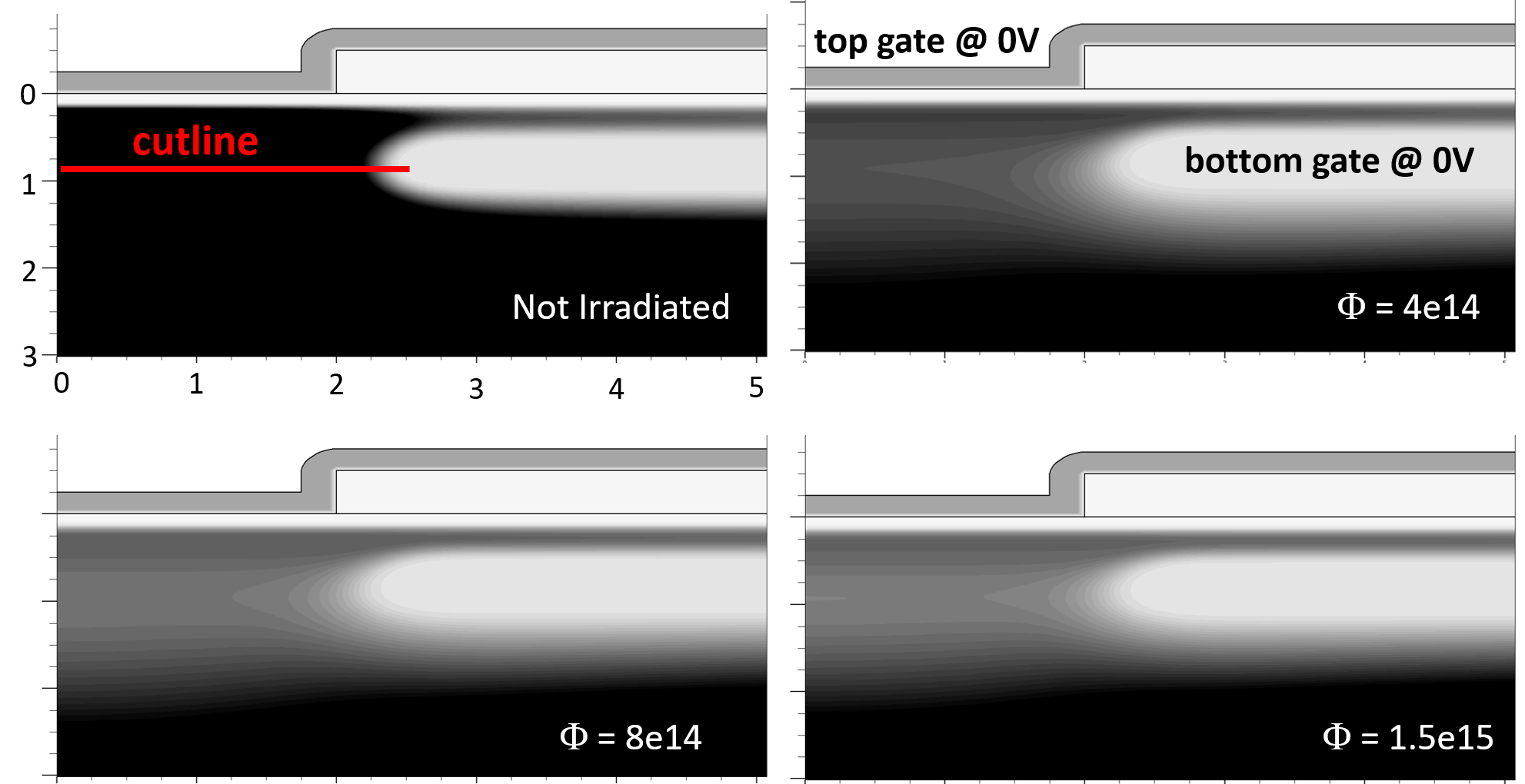}
\caption{\label{fig:potentialirradiated} Simulated electrostatic potential in the region of the gap, for different irradiation fluences. Drain voltage = - 100~V. The white colour maps equipotential regions at +0.5~V; black at -1~V. Potential, ionized donor trap density and charge density along the shown cutline are plotted in Figure~\ref{fig:cutlines}. The cutline has been drawn in the region where the potential has a maximum. }
\end{figure}

To visualize these effects,  the 2-dimensional map of the electrostatic potential in the gap region for the different irradiation fluences is plotted in Figure~\ref{fig:potentialirradiated}, for ${\rm V_{gate} = 0~V}$ and ${\rm V_{drain} = -100~V}$. In the not-irradiated case, the device is already in saturation: high-value (here and in the following the absolute value is understood) equipotential lines enter the gap region, reach the channel end and pinch it off. This is not the case for irradiated devices, where potentials closer to zero are present. The low voltage present in the case of a fluence of ${\rm 4 \cdot 10^{14}~ cm^{-2}}$ is  barely effective in closing the channel, while at the higher irradiation levels this voltage is not large enough to close the channel: lower potentials exist in the gap region  with the consequence that higher drain voltages must be applied to have values high enough to close the channel. Moreover, a potential barrier exists between the channel and the substrate across the gap region, that the holes have to cross to pass from channel to drain. This potential barrier is lowered by the application of higher drain voltages. 

\begin{figure}
\centering 
\includegraphics[width=.65\textwidth]{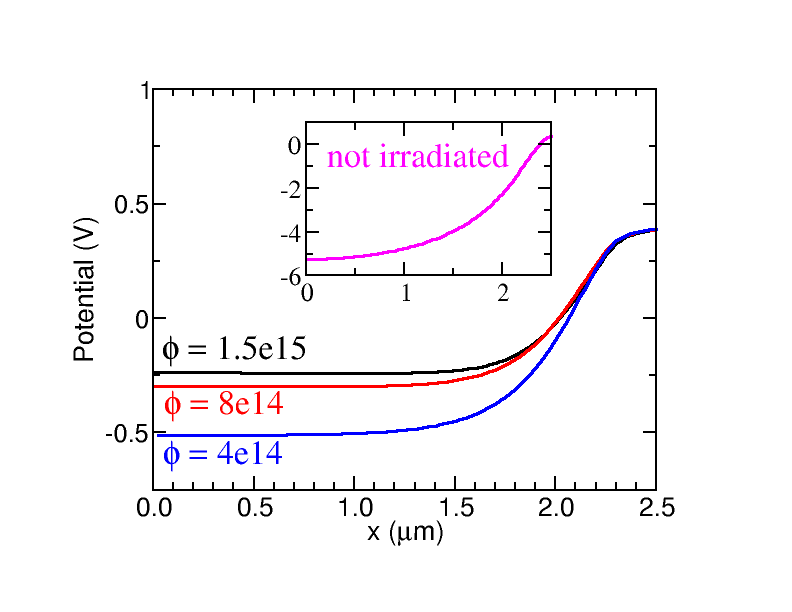}
\includegraphics[width=.65\textwidth]{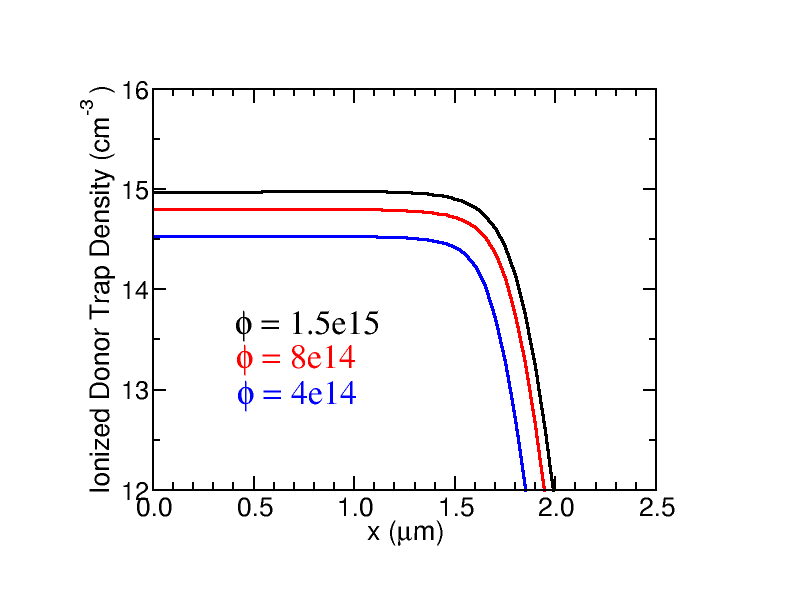}
\includegraphics[width=.65\textwidth]{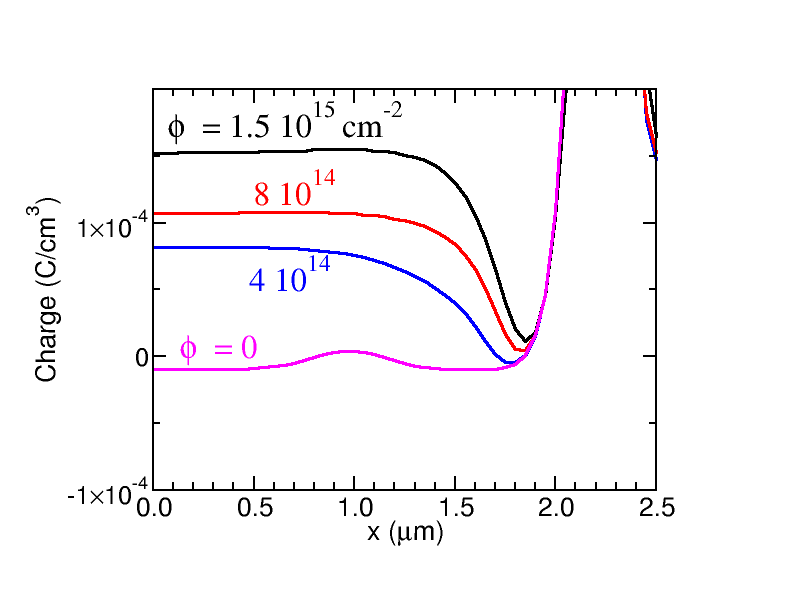}
\caption{\label{fig:cutlines} Top, cutlines of the electrostatic potential along the  line shown in Figure~\ref{fig:potentialirradiated}, for different values of the fluence $\Phi$, in ${\rm n_{eq}/cm^2}$. The tip of the bottom gate is at $x=2~\mu$m. Middle, density of the ionized donor traps along the same cutline, log scale. Bottom, charge density. Drain voltage = - 100~V.}
\end{figure}

Let's also consider the horizontal cutline along the lowest potential of Figure~\ref{fig:potentialirradiated} and shown in Figure~\ref{fig:cutlines}: the higher the irradiation level the lower the potential value in the gap region, while high voltages are needed in this region to close the channel. Also, a lower (closer to zero) potential means  a higher value of the potential barrier for the holes to cross before going to the substrate. This is correlated with the the positive charge created by the  ionized donor trap density  (also plotted in Figure~\ref{fig:cutlines}).

Finally, from Figure ~\ref{fig:simulations} it can be noted that, for a same gate voltage, the drain current in saturation is independent from the fluence, dramatically differing from measurements. This, as said above, is due to the fact that the Perugia model does not account for any acceptor removal phenomena.

\section{Conclusions}
\label{sec5}

HV vertical silicon JFETs, developed and fabricated at BNL,  have been irradiated up to a fluence of $\phi = 1.5 \cdot 10^{15}~ {\rm  n_{eq}cm^{-2}}$, which is the maximum fluence that the switch  for the HV-Mux   of the ATLAS ITk at the HL-LHC is expected to be exposed to. A static characterization at the probe station has been performed to assess their radiation hardness.  The chosen JFET, which belonged to the very first fabrication and must be seen as prototypes,  featured a breakdown voltage below the required by the HV-Mux specs; however it is found to be unaffected by the irradiation level. Most notably,  by measuring the output characteristics (i.e. drain currents vs drain voltage for different gate voltages), we can see how the drain saturation voltage is strongly dependent on the fluence:  at low drain voltages the drain currents stay on a common envelope, as in a standard JFET, while they depart from it at increasingly higher voltages as the fluence is increased. Numerical TCAD simulations explain the phenomenon to be due to a positive charge build-up, caused by the empty ionized donor traps induced by the irradiation. Consequently, a potential barrier is created for the holes (of the source-to-drain current) to cross. To be used as a switch in the HV-Mux, in the ON state the  current has to be in the order  of a few mA which requires for this family of JFETs an operational voltage of ${\rm V_{drain}>200~V}$, causing a large power dissipation within the device. Breakdown voltage, ON and OFF currents are strongly inter-correlated and depend on many  factors (size, channel doping concentration, geometry) as well as on the process flow: an optimization and an improvement of these prototype JFETs that push their performance towards in-spec HV-Mux switch  is certainly possible. However, a not negligible power dissipation within the device is always expected, particularly after irradiation; due to the outstanding performance of GaN JFETs, silicon HV JFETs are not going to be further considered for this application.

\acknowledgments
The team at the TRIGA nuclear reactor at JSI (Ljubljana, Slovenia)   is deeply acknowledged for having performed  the irradiation. This material is based upon work supported by the U.S. Department of Energy under grant DE-SC0012704. This research used resources of the Center for Functional Nanomaterials, which is a U.S. DOE Office of Science Facility, at Brookhaven National Laboratory under Contract No. DE-SC0012704.

\bibliographystyle{report}
\bibliography{biblio}{}


\end{document}